# Effects of Renewable Energy Sources on Day-Ahead Electricity Markets

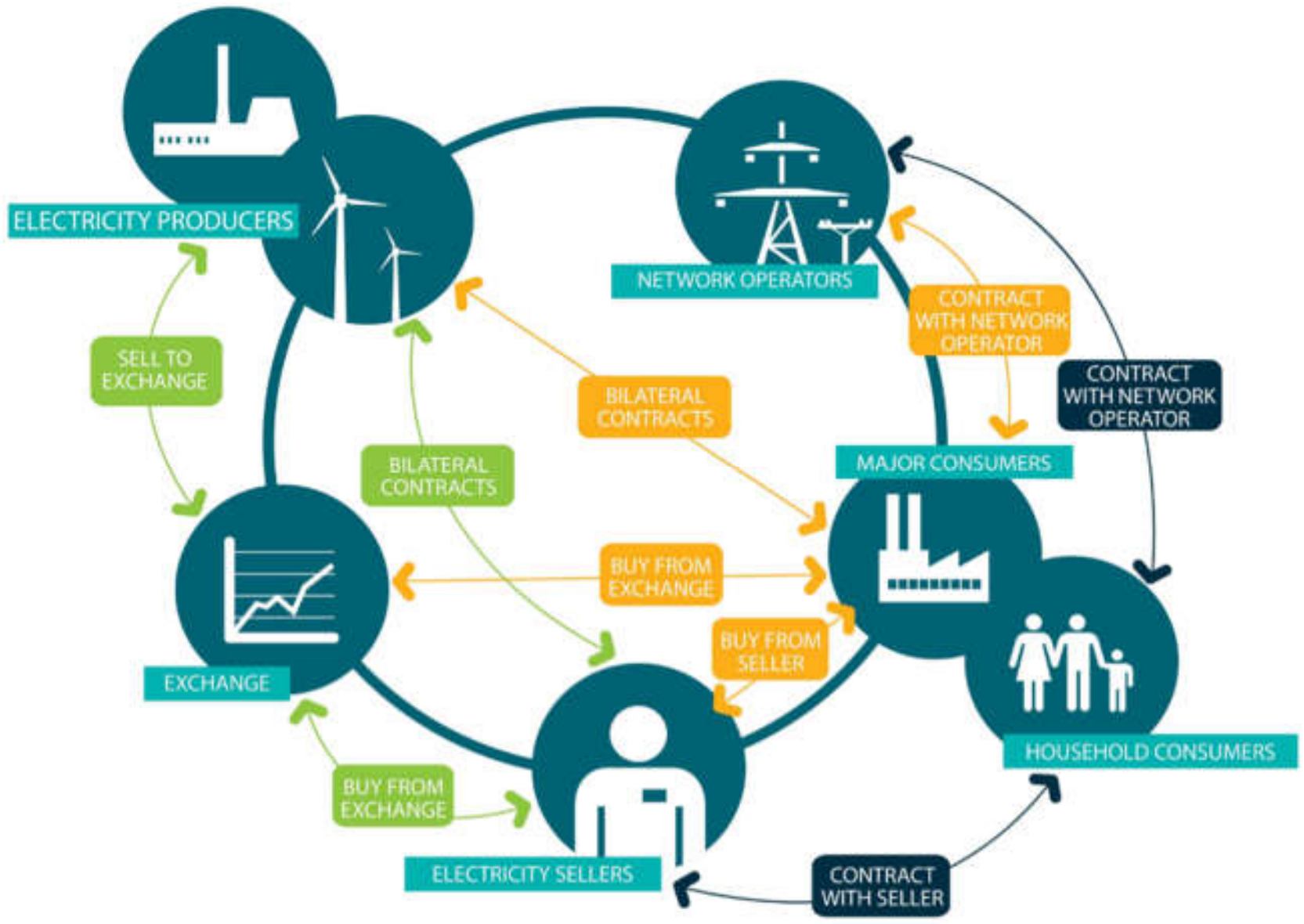

Author

Muhammad Hamad Zaheer

# Contents

# Abstract

The price of solar panels is decreasing rapidly and governments around the world are investing in solar power plants to increase the portion of solar energy in the smart grid. The price of electricity produced by solar power plants is much cheaper than conventional thermal power plants and innovation in solar cells is further increasing the divide. Consequently, setting up solar power plants will decrease electricity spot prices on average. However, the topology of the power grid must be considered while setting up solar power plants so that full benefits of cheap solar power can be gained by all consumers.

In this paper we will calculate day-ahead spot prices using conventional generators. Then one of the generators will be replaced by a photovoltaic generator and effects on day-ahead market will be analyzed. Total cost of electricity generated at each hour in a day using conventional power plants and solar power plants will be compared. The effects of line congestion on the benefits of solar power to end customers will be analyzed by comparing nodal prices.

The optimization problems related to generator scheduling and day-ahead markets will be solved using Dynamic Programming, Mixed Integer Linear Programming (MILP) and Karush-Kuhn-Tucker (KKT) method for convex optimization.

# 1 Problem Statement

In this report we will show that cheap electricity due to solar power plants decrease the day-ahead nodal prices. However, due to line congestion some nodal prices remain the same or show little decrease resulting in some customers not enjoying the full benefits of cheaper solar power. Therefore, solar power plants should be installed strategically taking into account the topology of the power grid.

The optimization problem of day-ahead markets is solved in two steps. First generation scheduling optimization problem is solved. The scheduling problem will be solved over a 24-hour period. We need a forecast of demand to solve the scheduling problem. Figure 1-1 shows a typical demand curve. A demand curve similar to this one will be used as forecasted demand for optimal scheduling of generators. Figure 1-2 [2] shows the forecasted demand curve that will be used in this report. The optimal scheduling problem gives the turn on and off status of generators at each hour and their output power.

After optimal scheduling problem we will solve day-ahead market problem and find day-ahead nodal prices at each hour in a day. Using the status of generators at each hour from the optimal scheduling step, the generator dispatch problem is solved at each hour. Then we will calculate the nodal prices at each hour using the inequality and equality constraints of the optimal dispatch problem.

The optimization problem of day-ahead market will be solved using different techniques. Dynamic programming and mixed integer linear programming will be used to solve the optimization problem. The optimal dispatch problem will be solved using logical methods, formula method and programming methods.

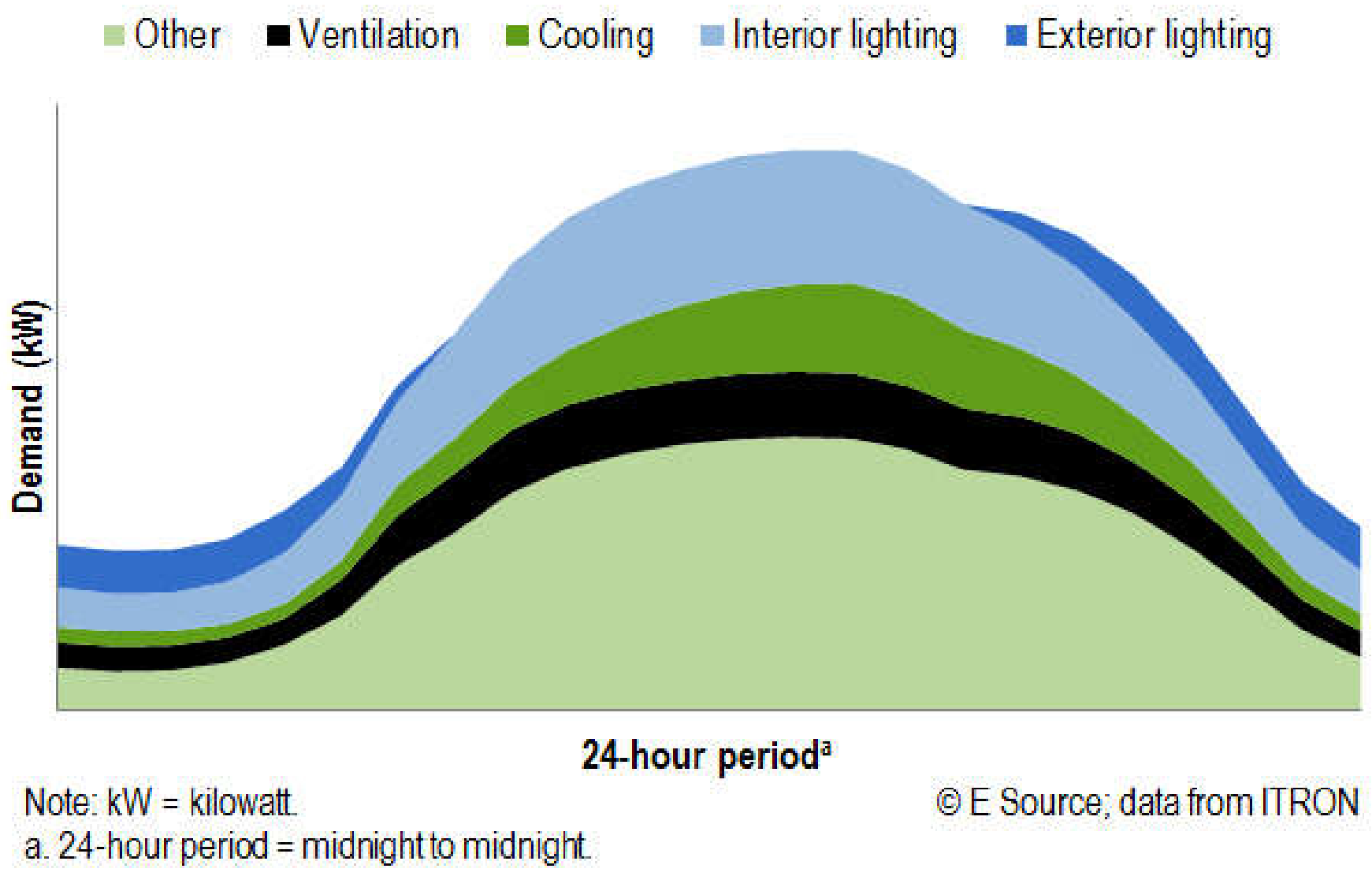

Figure 1-1Typical Demand Curve

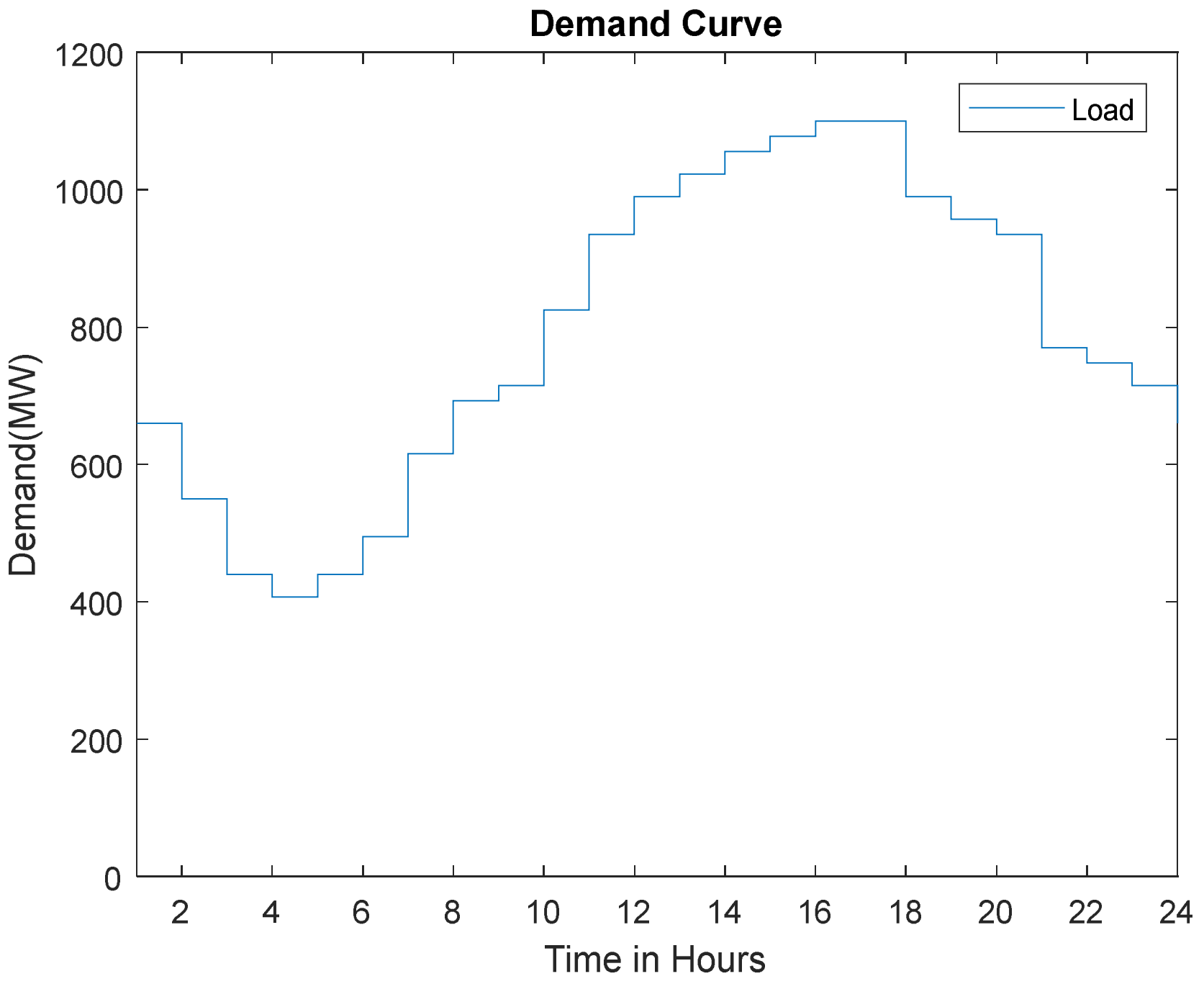

Figure 1-2 Demand Curve for Generator Scheduling

## 1.1 System Model

A 7-Bus (node) System will be considered to evaluate the effects of solar energy on day-ahead electricity prices. Figure 1 shows the one-line diagram of the system. The system consists of three generators (G1, G2, and G3). The data for power generating units, loads and transmission lines is given in Table 1, Table 2 and Table 3 respectively.

A solar panel based generator Gx is connected in parallel to G2. We will first calculate day-ahead prices using conventional generator G2 and then replace it with Gx and discuss the effects of solar energy on day-ahead markets.

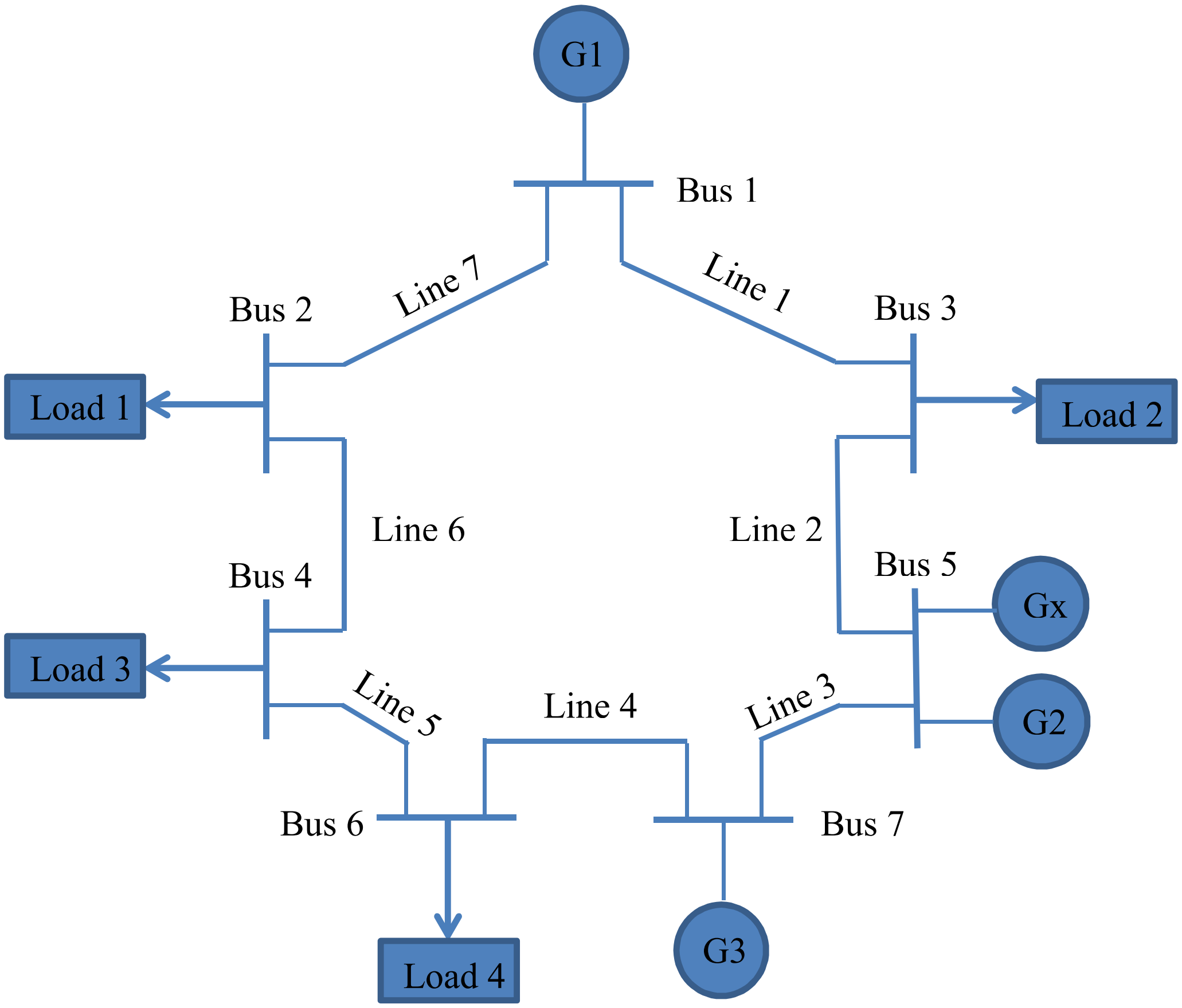

Figure 1-3 One-line diagram of 7 Bus System

| Unit | At bus | Pmin(MW) | Pmax(MW) | Startup Cost($) | Cost ($/MWh) |
|---|---|---|---|---|---|
| **G1** | 1 | 20 | 500 | 500 | 10 |
| **G2** | 5 | 10 | 500 | 300 | 20 |
| **G3** | 7 | 0 | 150 | 50 | 50 |
| **Gx (Solar)** | 5 | 0 | 500 | 0 | 5 |

Table 1-1 Generator data

| Unit | At bus | P(MW) |
|---|---|---|
| **Load 1** | 2 | 300 |
| **Load 2** | 3 | 200 |
| **Load 3** | 4 | 300 |
| **Load 4** | 6 | 300 |

Table 1-2Load data

| Line | From bus | To bus | x(p.u.) | $b_{ij}$ | Flow limit(MW) |
|---|---|---|---|---|---|
| **1** | 1 | 3 | 0.05 | 20 | 800 |
| **2** | 3 | 5 | 0.05 | 50 | 300 |
| **3** | 5 | 7 | 0.05 | 100 | 300 |
| **4** | 7 | 6 | 0.05 | 10 | 800 |
| **5** | 6 | 4 | 0.05 | 100 | 800 |
| **6** | 4 | 2 | 0.05 | 50 | 500 |
| **7** | 2 | 1 | 0.05 | 20 | 500 |

Table 1-3Transmission line data

# 1.2 System Parameters

Based on the system model, the admittance matrix is given as:

$$B = \begin{bmatrix} 40 & -20 & -20 & 0 & 0 & 0 & 0 \\ -20 & 70 & 0 & -50 & 0 & 0 & 0 \\ -20 & 0 & 70 & 0 & -50 & 0 & 0 \\ 0 & -50 & 0 & 150 & 0 & -100 & 0 \\ 0 & 0 & -50 & 0 & 150 & 0 & -100 \\ 0 & 0 & 0 & -100 & 0 & 110 & -10 \\ 0 & 0 & 0 & 0 & -100 & -10 & 110 \end{bmatrix}$$

Choosing bus 7 as reference node we get:

$$B' = \begin{bmatrix} 40 & -20 & -20 & 0 & 0 & 0 \\ -20 & 70 & 0 & -50 & 0 & 0 \\ -20 & 0 & 70 & 0 & -50 & 0 \\ 0 & -50 & 0 & 150 & 0 & -100 \\ 0 & 0 & -50 & 0 & 150 & 0 \\ 0 & 0 & 0 & -100 & 0 & 110 \end{bmatrix}$$

The matrix $X$ and $T$ are given as:

$$X = \begin{bmatrix} -20 & 20 & 0 & 0 & 0 & 0 & 0 \\ -50 & 0 & 50 & 0 & 0 & 0 & 0 \\ 0 & 100 & 0 & -100 & 0 & 0 & 0 \\ 0 & 0 & 10 & 0 & -10 & 0 & 0 \\ 0 & 0 & 0 & 100 & 0 & -100 & 0 \\ 0 & 0 & 0 & 0 & 50 & 0 & -50 \\ 0 & 0 & 0 & 0 & 0 & 20 & -20 \end{bmatrix}$$

$$\boldsymbol{T} = X \begin{bmatrix} B'^{-1} & 0 \\ 0 & 0 \end{bmatrix}$$

# 2 Generation Scheduling

## 2.1 Optimal Generator Scheduling

We are interested in finding day-ahead spot prices. For that we first need to solve the optimal scheduling problem. Consider the following line flow constrains:

$$\overline{\boldsymbol{F}}^T = \begin{bmatrix} 800 & 300 & 300 & 800 & 800 & 500 & 500 \end{bmatrix}^T$$

Define:

$$\boldsymbol{P}_G^t - \boldsymbol{P}_D^t = \begin{bmatrix} P_{G1}^t & -P_{L1}^t & -P_{L2}^t & -P_{L3}^t & P_{G2}^t & -P_{L4}^t & P_{G3}^t \end{bmatrix}^T$$

The optimization problem for solving the scheduling problem is [3]:

$$\underset{P_G, U}{Min} \sum_{i=1}^{NG} \sum_{j=1}^{24} \left[ U_i^t c_i P_{Gi}^t + S_i^t U_i^t \left(1 - U_i^{t-1}\right) \right]$$

$$\text{S.T.} \quad \boldsymbol{e}^T \left( \boldsymbol{P}_G^t - \boldsymbol{P}_D^t \right) = 0, (t = 1, 2, \cdots, 24)$$

$$\boldsymbol{T} \left( \boldsymbol{P}_G^t - \boldsymbol{P}_D^t \right) \le \overline{\boldsymbol{F}}, (t = 1, 2, \cdots, 24)$$

$$U_i^t \underline{P}_{Gi}^t \le P_{Gi}^t \le U_i^t \overline{P}_{Gi}^t, (t = 1, 2, \cdots, 24)(i = 1, 2, \cdots, NG)$$

$$U_i^t \in \{0, 1\}, (t = 1, 2, \cdots, 24)(i = 1, 2, \cdots, NG)$$

## 2.2 Solving Generator Scheduling Optimization Problem

The unit commitment problem can't be solved by employing economic dispatch at each hour. This is because of startup costs and ramp rates of individual generators which create a link between different time periods. Moreover, the power generated by each generator is either 0 or bounded between non-zero bounds. This is represented mathematically by multiplying lower and upper limits by binary variables, $U_i^t$ . Therefore, because of the presence of binary variables linear programming techniques can't be used to solve generator scheduling problem. Dynamic programming or mixed integer linear programming is employed to solve the optimal generator scheduling problem.

## 2.2.1 Method 1: Dynamic Programming

In dynamic programming we divide our problem to a number of sub problems like divide and conquer algorithms and find the optimal solutions of each sub problem. However, unlike divide and conquer algorithms the sub problems can overlap each other.

Table 2-2 shows different generator combinations that can be used during each time interval. Figure 2-1 shows all the possible states at each time interval. The feasible states are the ones in which the generators that are turned on can provide the required load. The feasible states are shown in light and dark blue. However, light blue states are not feasible when line constraints are used. Therefore, only dark blue states are feasible. The arrows show state transitions. We can follow the arrows to find all feasible transitions.

We are interested in finding the sequence of state transitions that is cheapest and hence optimal. The amount at the top of the circles shows the total cost of generation at that time interval. The orange colored amounts at the arrows are the start-up cost of generators. The optimal solution is the sequence 1-2-4-7 shown in blue color in Table 2-1.

The 24-hr scheduling problem is solved using the same algorithm in Matlab. The results are shown in 2.3.

| State Sequences | Cost |
|---|---|
| 1-2-3-6 | $22,870 |
| 1-2-3-7 | $22,770 |
| 1-2-3-8 | $26,840 |
| 1-2-4-6 | $22,870 |
| 1-2-4-7 | $22,670 |
| 1-2-4-8 | $26,840 |
| 1-2-5-6 | $27,670 |
| 1-2-5-7 | $27,570 |
| 1-2-5-8 | $31,140 |

Table 2-1State Sequences and total cost

| Combinations G1 | G2 | G3 | Pmin (MW) | Pmax (MW) |
|---|---|---|---|---|
| 1 | 1 | 1 | 30 | 1150 |
| 1 | 1 | 0 | 30 | 1000 |
| 1 | 0 | 1 | 20 | 650 |
| 1 | 0 | 0 | 20 | 500 |
| 0 | 1 | 1 | 10 | 650 |
| 0 | 1 | 0 | 10 | 500 |
| 0 | 0 | 1 | 0 | 150 |
| 0 | 0 | 0 | 0 | 0 |

Table 2-2Generator Combinations

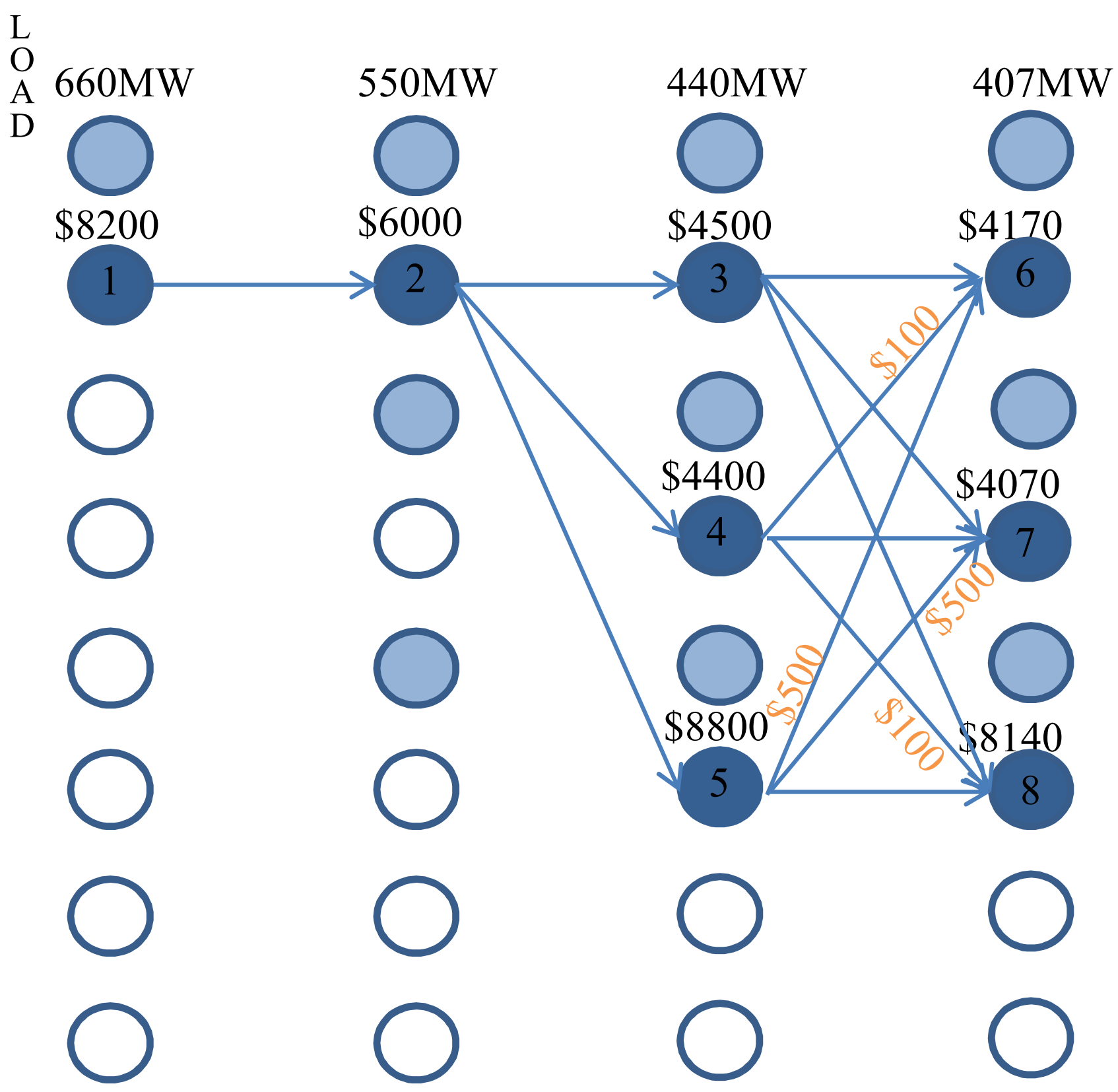

Figure 2-1Possible State Transitions

### 2.2.2 Method 2: Mixed Integer Linear Programming

Mixed integer liner programming (MILP) is used to solve optimization problems which contain integer variables instead of real number variables. Simplex method is mostly used for linear programming. Revised simplex method which is based on the simplex method is used to solve MILP. Several libraries are available to solve MILP. We used YALMIP and Matlab to solve to MILP for generator scheduling [4]. Code is given is Appendix.

## 2.3 Results of Optimal Generator Scheduling

Figure 2-2 and Figure 2-3 represents the generator scheduling over a 24 hour period. In Figure 2-2 no renewable source is used. In Figure 2-3 thermal generator G2 is replaced with a solar photovoltaic based generator Gx. Because solar energy is cheaper it replaces expensive sources to provide the base load. The expensive generators G1 and G3 are scheduled only during peak hour and overall cost of electricity generation will be reduced.

The optimal generator scheduling also provides the turn on and off status of generators as each hour. Figure 2-4 and Figure 2-5 shows the status of generators using convention generator G2 and solar generator Gx respectively. This information will be used in the next section, 3 to solve the day-ahead market problem and calculate spot nodal prices at each hour.

Figure 2-6 and Figure 2-7 shows the social production cost (cost of total power generated) at each hour. The total social production cost has reduced when solar generator is used. Moreover, the peak cost at peak hours has also reduced. Therefore, renewable energy sources not only decrease the total cost but also peak cost of electricity and can make electricity a lot cheaper for consumers and industry.

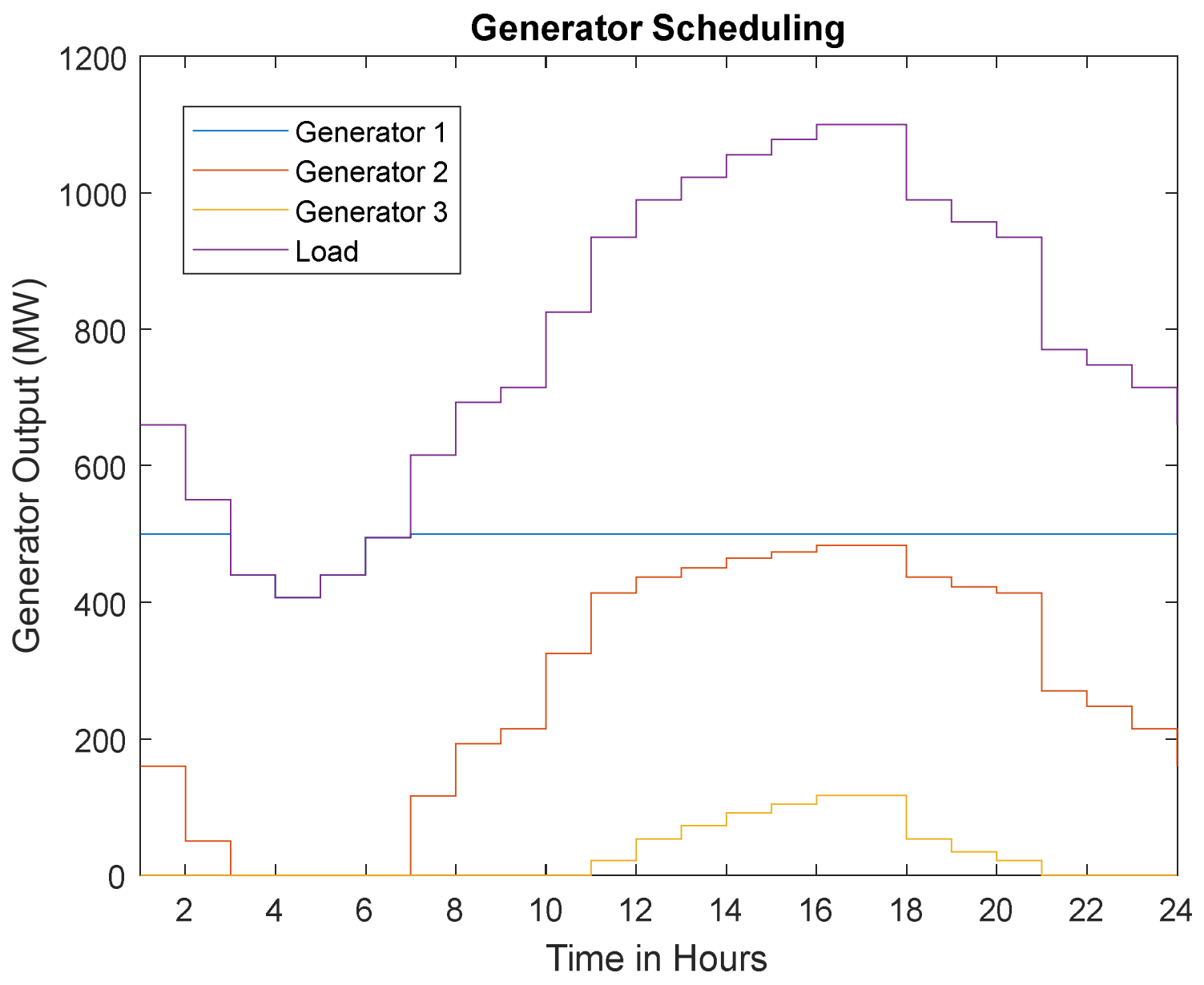

Figure 2-2Generator Scheduling over a 24 hr period

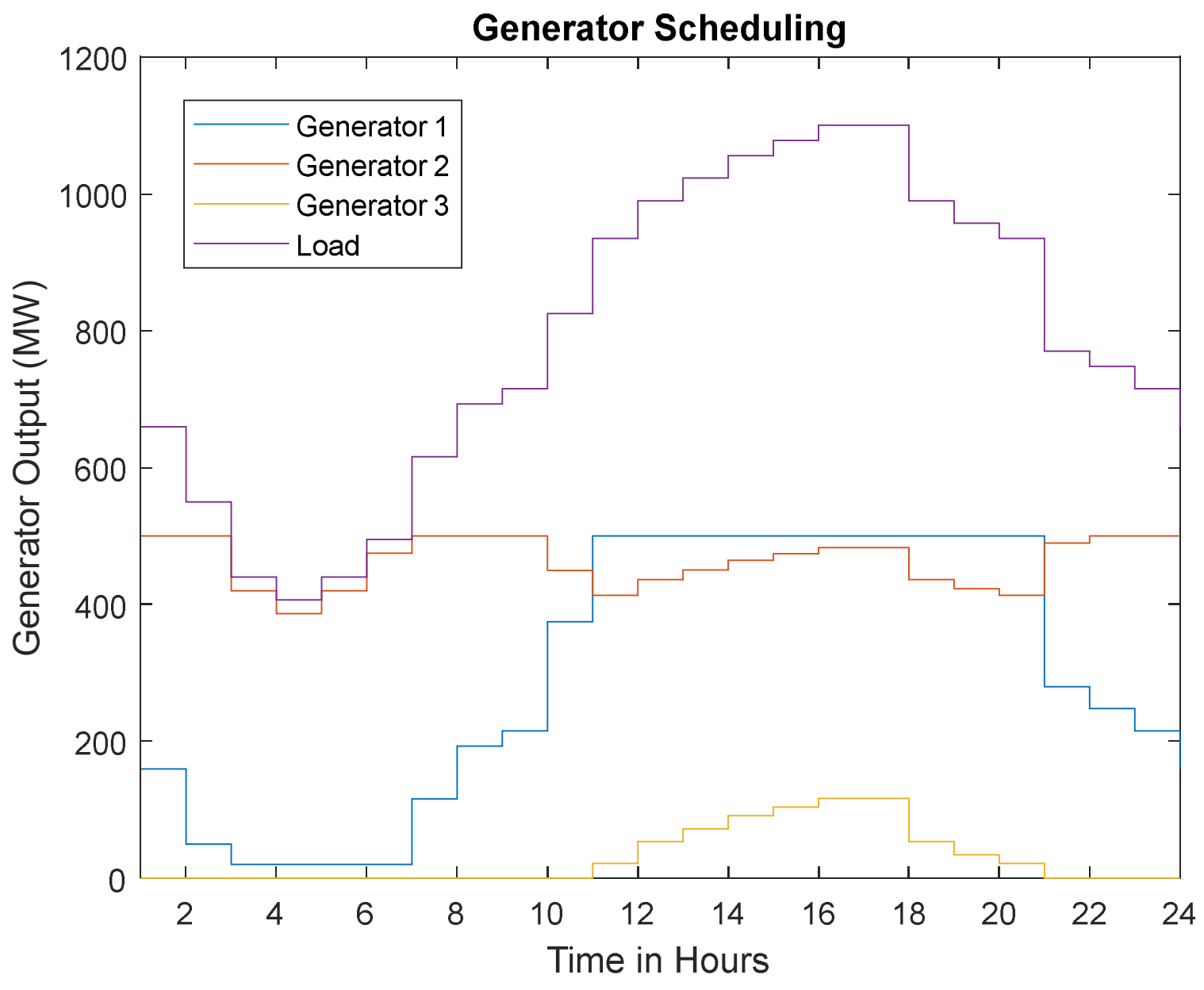

Figure 2-3Generator Scheduling with G2 replaced with solar power plant Gx

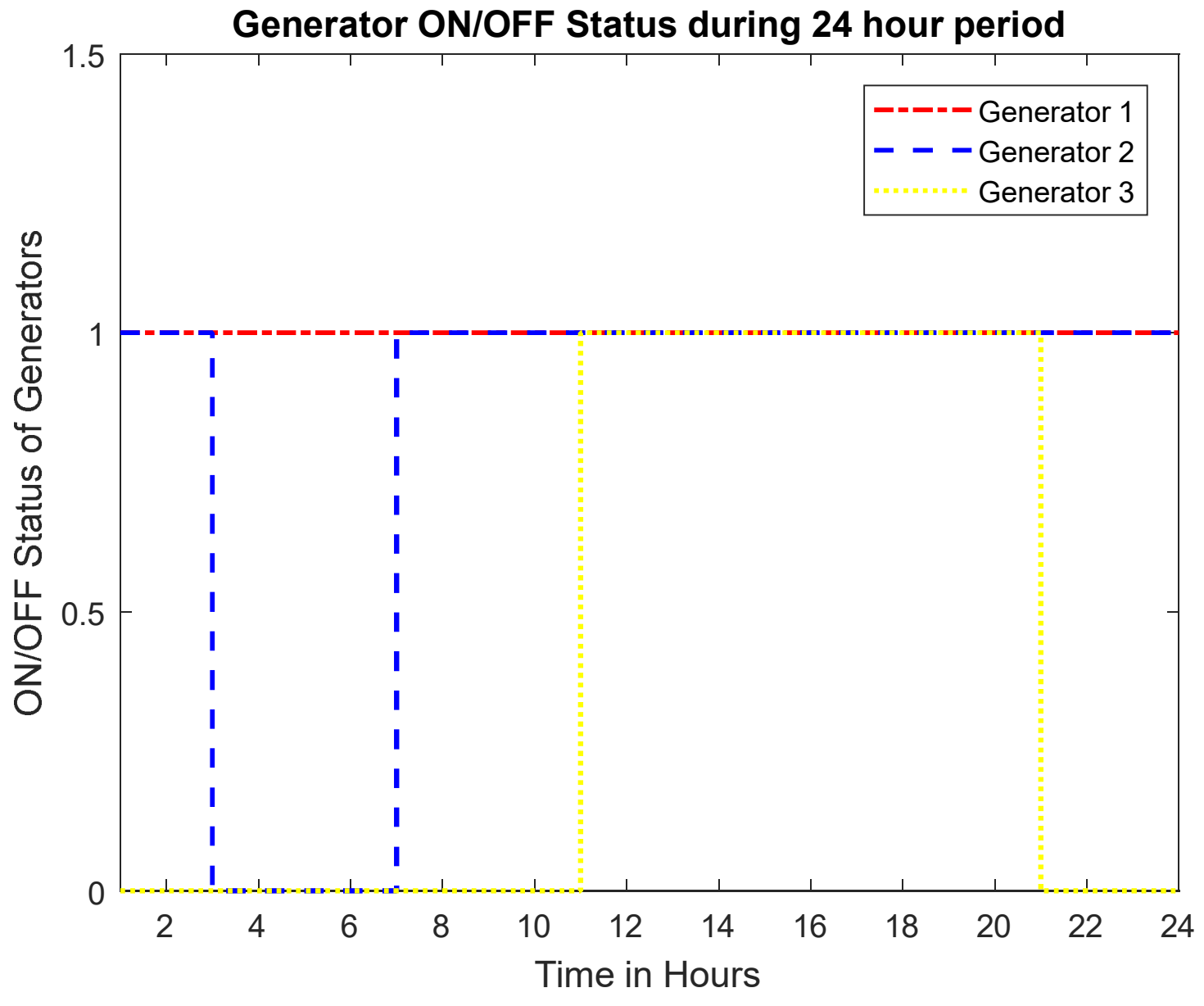

Figure 2-4 Generator Status during 24 hr period

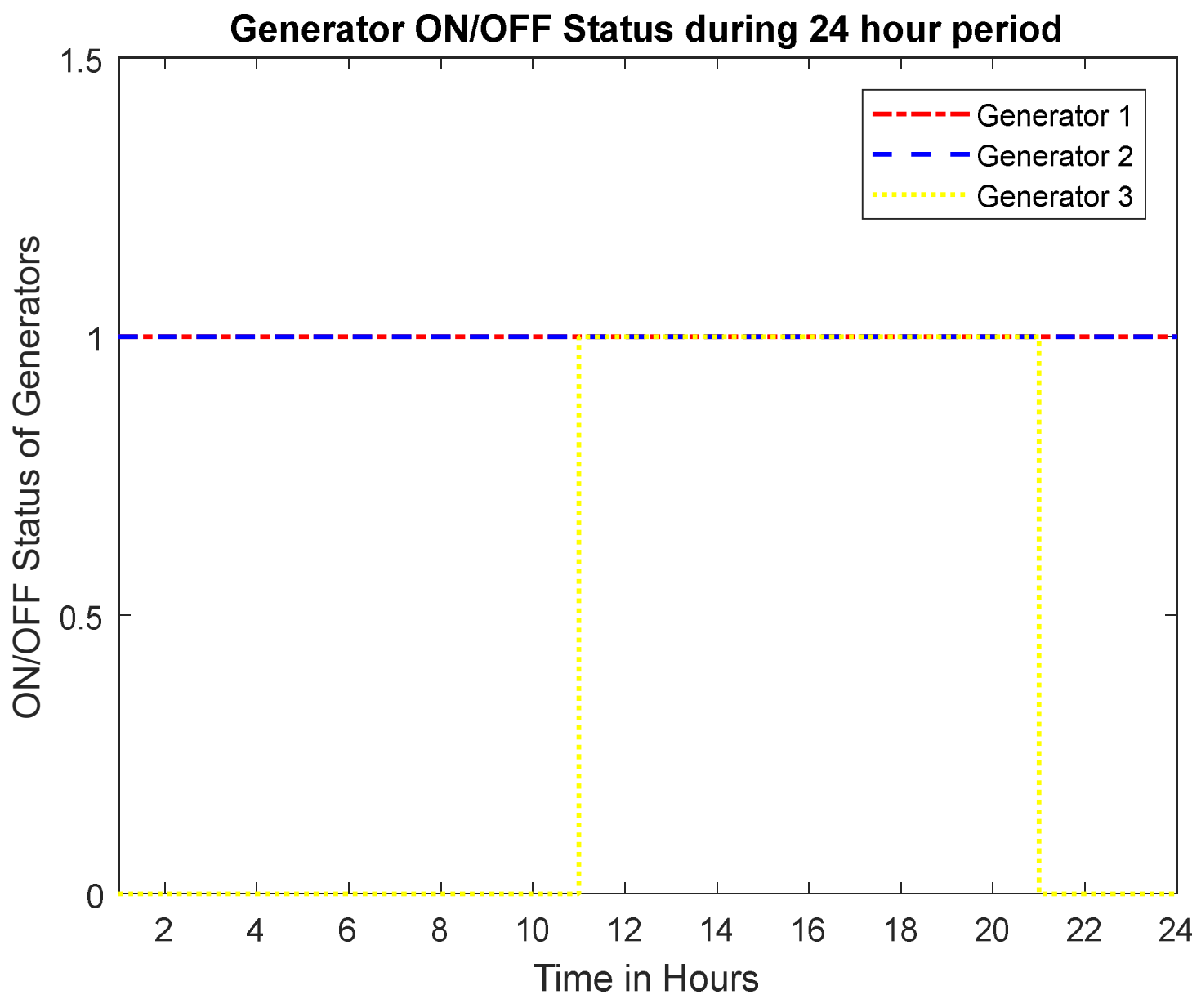

Figure 2-5 Generator Status using solar power plant Gx

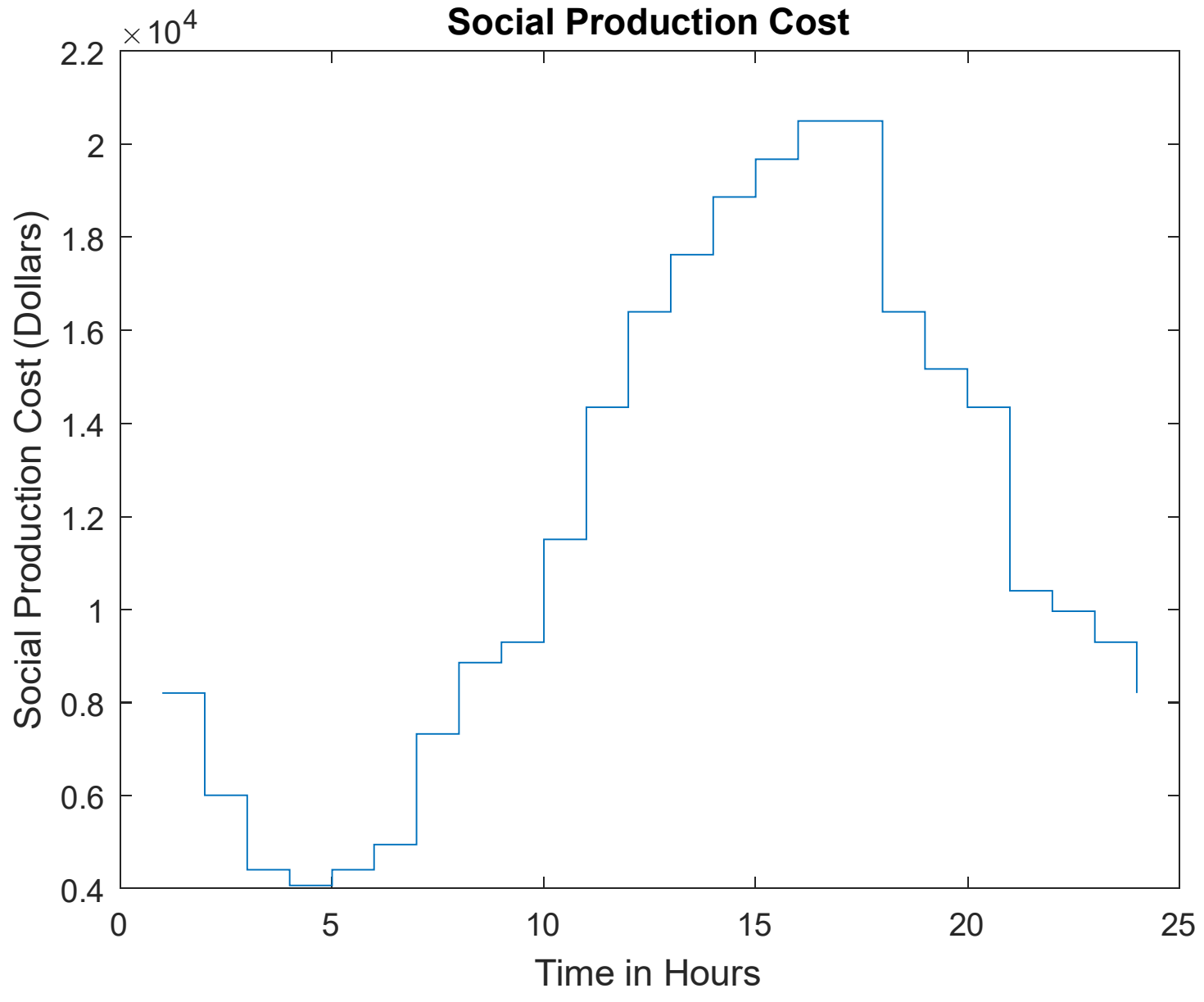

Figure 2-6 Social Production cost without solar power

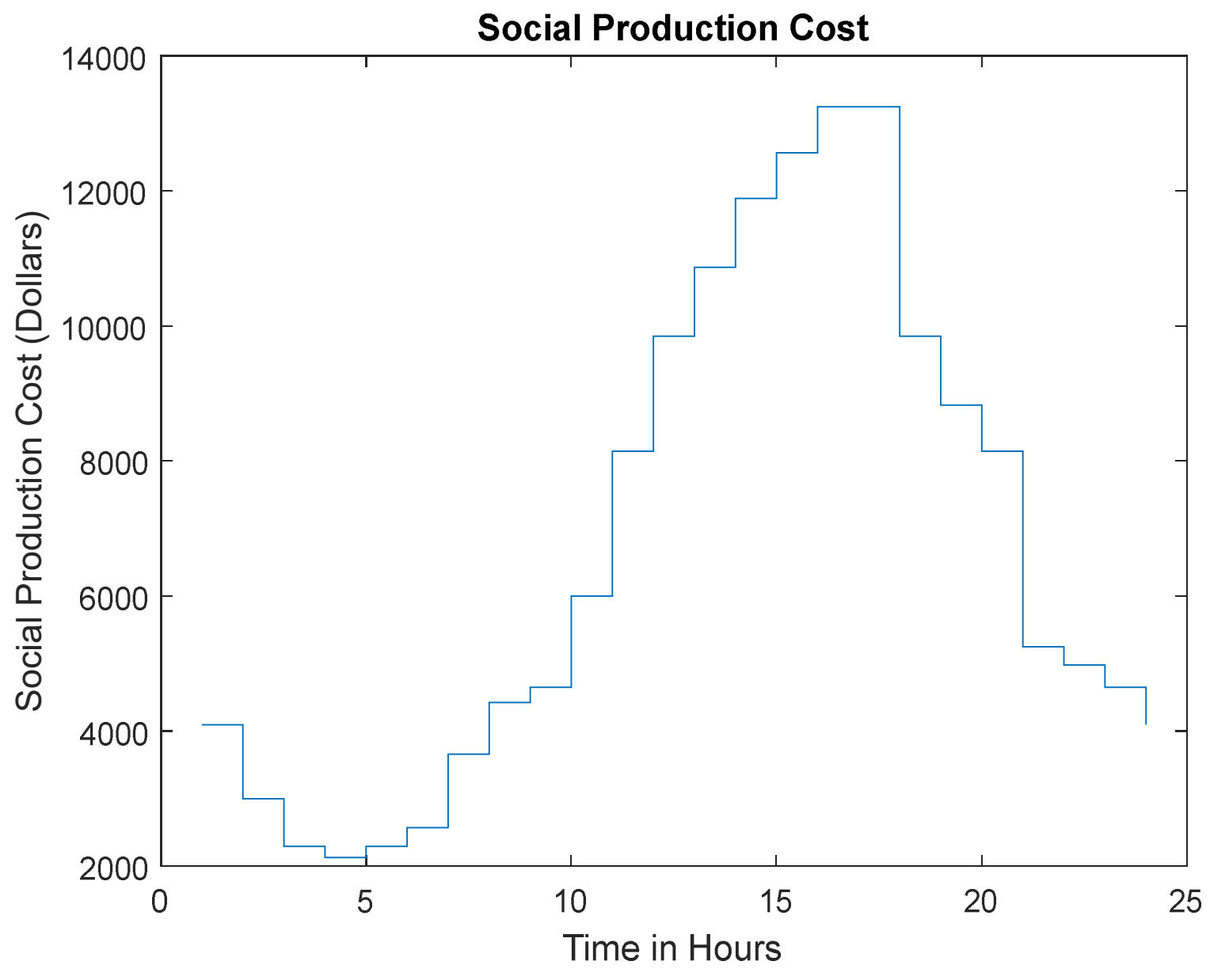

Figure 2-7 Social Production cost with solar power

# 3 Day-Ahead Markets

After solving the optimization problem of generation scheduling we get the on and off status of all generators for all time periods. We will use this information to solve day-ahead market problem for hourly nodal prices.

Let $I$ be the set of all generators which are operational at time $t$. We solve the following optimization problem to calculate nodal prices at each time $t$ [3].

$$\min_{P_G,U} \sum_{i\in I} c_i P_{Gi}^t$$

$$\text{S.T.} \quad \boldsymbol{e}^T\left(\boldsymbol{P}_G^t - \boldsymbol{P}_D^t\right) = 0, (t = 1, 2, \cdots, 24)$$

$$\boldsymbol{T}\left(\boldsymbol{P}_G^t - \boldsymbol{P}_D^t\right) \le \overline{\boldsymbol{F}}, (t = 1, 2, \cdots, 24)$$

$$\underline{P}_{Gi}^t \le P_{Gi}^t \le \overline{P}_{Gi}^t, (t = 1, 2, \cdots, 24)(i \in I)$$

## 3.1 Solving Day-Ahead Electricity Markets Problem

### 3.1.1 Method 1: Formula Method

The total demand at time $t$ is given by the sum of individual loads at $t$.

$$P_D^t = P_{L1}^t + P_{L2}^t + P_{L3}^t + P_{L4}^t$$

The generator power constraints are:

$$20 \le P_{G1}^t \le 500$$
$$10 \le P_{G2}^t \le 500$$
$$0 \le P_{G3}^t \le 150$$

The power balance equation is:

$$P_{G1}^t + P_{G2}^t + P_{G3}^t = P_D^t$$

The Lagrangian is given by:

$$\Gamma = 10P_{G1}^t + 20P_{G2}^t + 50P_{G3}^t + \lambda(P_{G1}^t + P_{G2}^t + P_{G3}^t - P_D^t) + \mu^T\left[\boldsymbol{T}\left(\boldsymbol{P}_G^t - \boldsymbol{P}_D^t\right) - \overline{\boldsymbol{F}}\right] + \left[\hat{\tau}_1(P_{G1}^t - 500) - \breve{\tau}_1(P_{G1}^t - 20)\right] + \left[\hat{\tau}_2(P_{G2}^t - 500) - \breve{\tau}_2(P_{G2}^t - 10)\right] + \left[\hat{\tau}_3(P_{G3}^t - 150) - \breve{\tau}_3 P_{G3}^t\right]$$

The optimal (Karush–Kuhn–Tucker) conditions are:

$$10 + \lambda + \sum_k \mu_k T_{k1} + \hat{\tau}_1 - \breve{\tau}_1 = 0$$
$$20 + \lambda + \sum_k \mu_k T_{k2} + \hat{\tau}_2 - \breve{\tau}_2 = 0$$
$$50 + \lambda + \sum_k \mu_k T_{k3} + \hat{\tau}_3 - \breve{\tau}_3 = 0$$

$$\mu^T\left[\boldsymbol{T}\left(\boldsymbol{P}_G^t - \boldsymbol{P}_D^t\right) - \overline{\boldsymbol{F}}\right] = 0$$

$$P_{G1}^t + P_{G2}^t + P_{G3}^t = P_D^t$$

$$\hat{\tau}_1(P_{G1}^t - 500) = 0,\ \ \breve{\tau}_1(10 - P_{G1}^t) = 0$$
$$\hat{\tau}_2(P_{G2}^t - 500) = 0,\ \ \breve{\tau}_2(20 - P_{G2}^t) = 0$$
$$\hat{\tau}_3(P_{G3}^t - 150) = 0,\ \ \breve{\tau}_3(-P_{G3}^t) = 0$$

Where,

$$\mu \ge 0$$

$$\hat{\tau}_1 \ge 0, \hat{\tau}_2 \ge 0, \hat{\tau}_3 \ge 0$$
$$\breve{\tau}_1 \ge 0, \breve{\tau}_2 \ge 0, \breve{\tau}_3 \ge 0$$

Nodal price of node i can be calculated by:

$$\rho_i = -\lambda - \sum_k \mu_k T_{ki}$$

We solve this problem for $t$=17, the time with maximum load.

| Generator Output | | | Nodal Prices $/MW | | | | | | |
|---|---|---|---|---|---|---|---|---|---|
| $P_{G1}$ | $P_{G2}$ | $P_{G3}$ | $\rho_1$ | $\rho_2$ | $\rho_3$ | $\rho_4$ | $\rho_5$ | $\rho_6$ | $\rho_7$ |
| 500 | 483.33 | 116.67 | 30 | 35 | 25 | 40 | 20 | 45 | 50 |

Table 3-1Optimal solution when load = 1100 MW (time=17)

Similarly we can solve the optimization problem to find hourly nodal prices for the 24 hour period using YALMIP and Matlab [4].

## 3.1.2 Method 2: Logical Method

The logical method [3] is used to get a logical understanding of the day-ahead market's optimization problem. According to the logical method [3] we first utilize the cheapest generator and take as much power as we can. However, if there are line constraints then we should also consider them and dispatch the generators in a way that flow limits are not violated.

**Step 1:**

First we will use the cheapest generator. The cheapest generator is G1 with a capacity of 500 MW and a cost of 10 $/MWh. The generator output should be increased until a line flow constraint is violated or the maximum generation capacity is reached. The flow rates at each line are indicated in green in Figure 3-1. The flow rates should be less than or equal to the maximum flow rate allowable to avoid damage to the line.

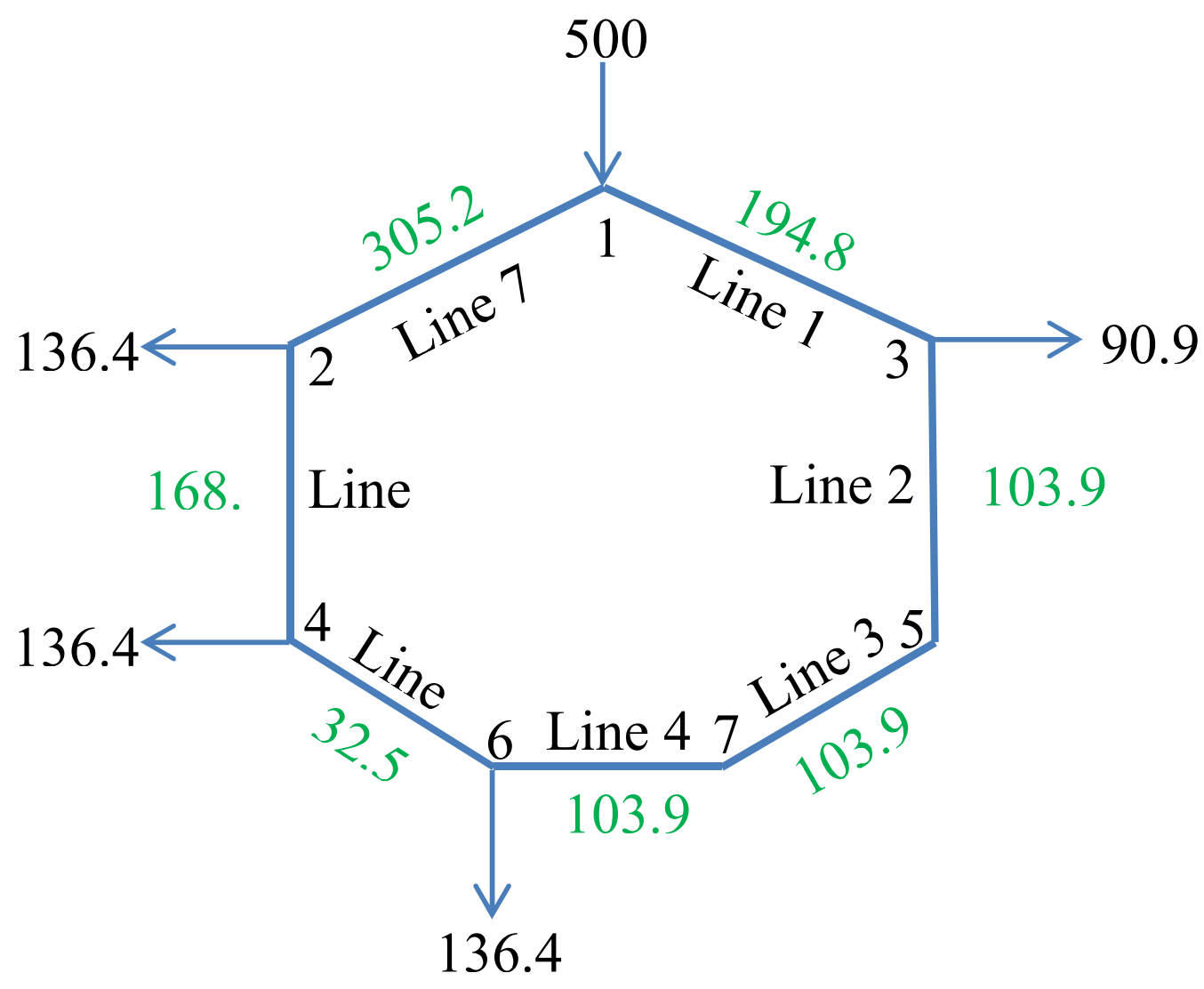

Figure 3-1 Generator 1 dispatched to 500MW

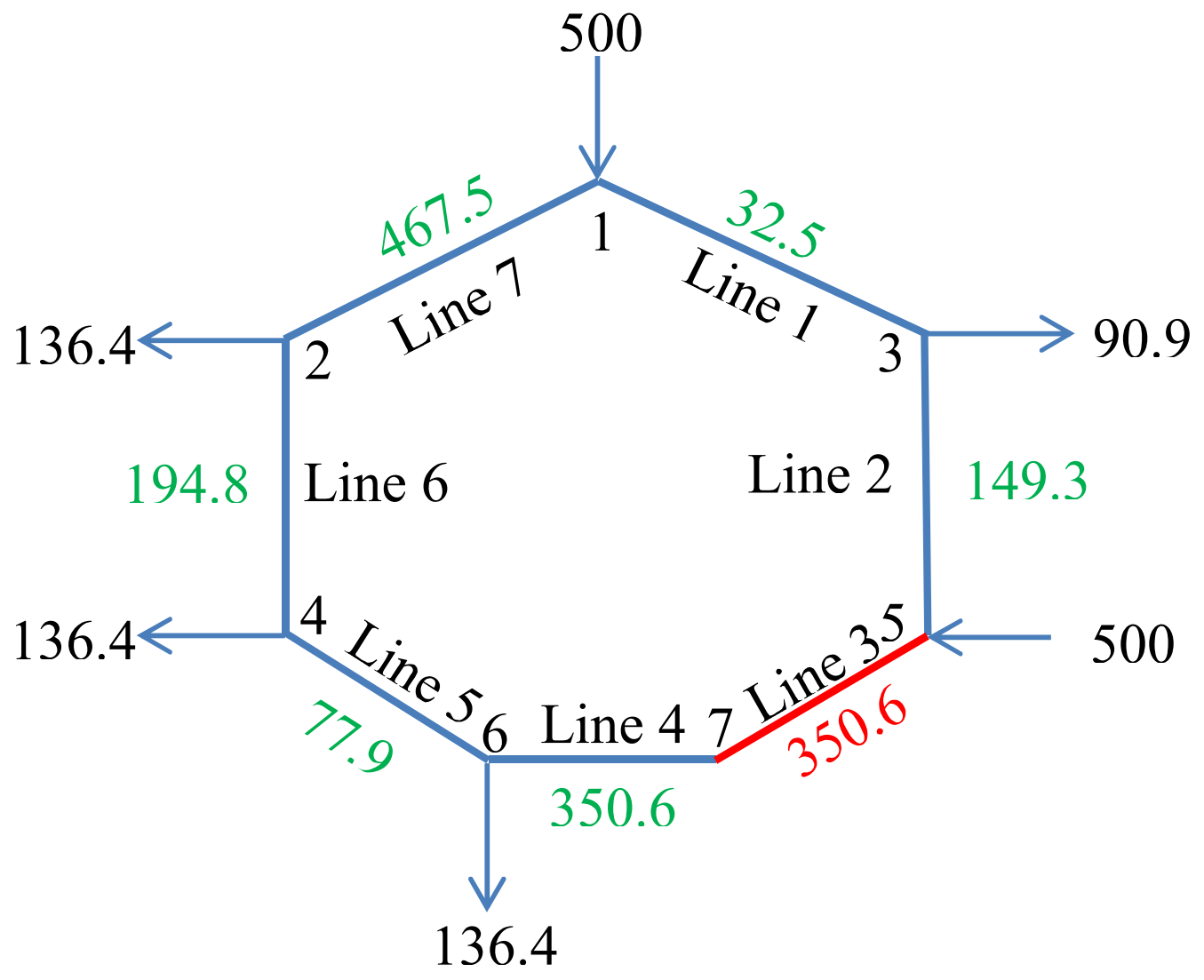

Figure 3-2 Generator 2 dispatched to 500MW (Line 3 overloaded)

**Step 2:**

Now we will dispatch the next cheapest generator. The next cheapest generator is G2 with a capacity of 500 MW and a cost of 20 $/MWh. Now we will use maximum power from this unit. The flow rates in red indicate a violation of the flow constraints.

In Figure 3-2 line 3 is overloaded. Therefore we have to decrease the output power of generator 2.

**Step 3:**

Now we decrease the output of generator 2 to 400MW as shown in Figure 3-3, so that no line flow constraints are violated.

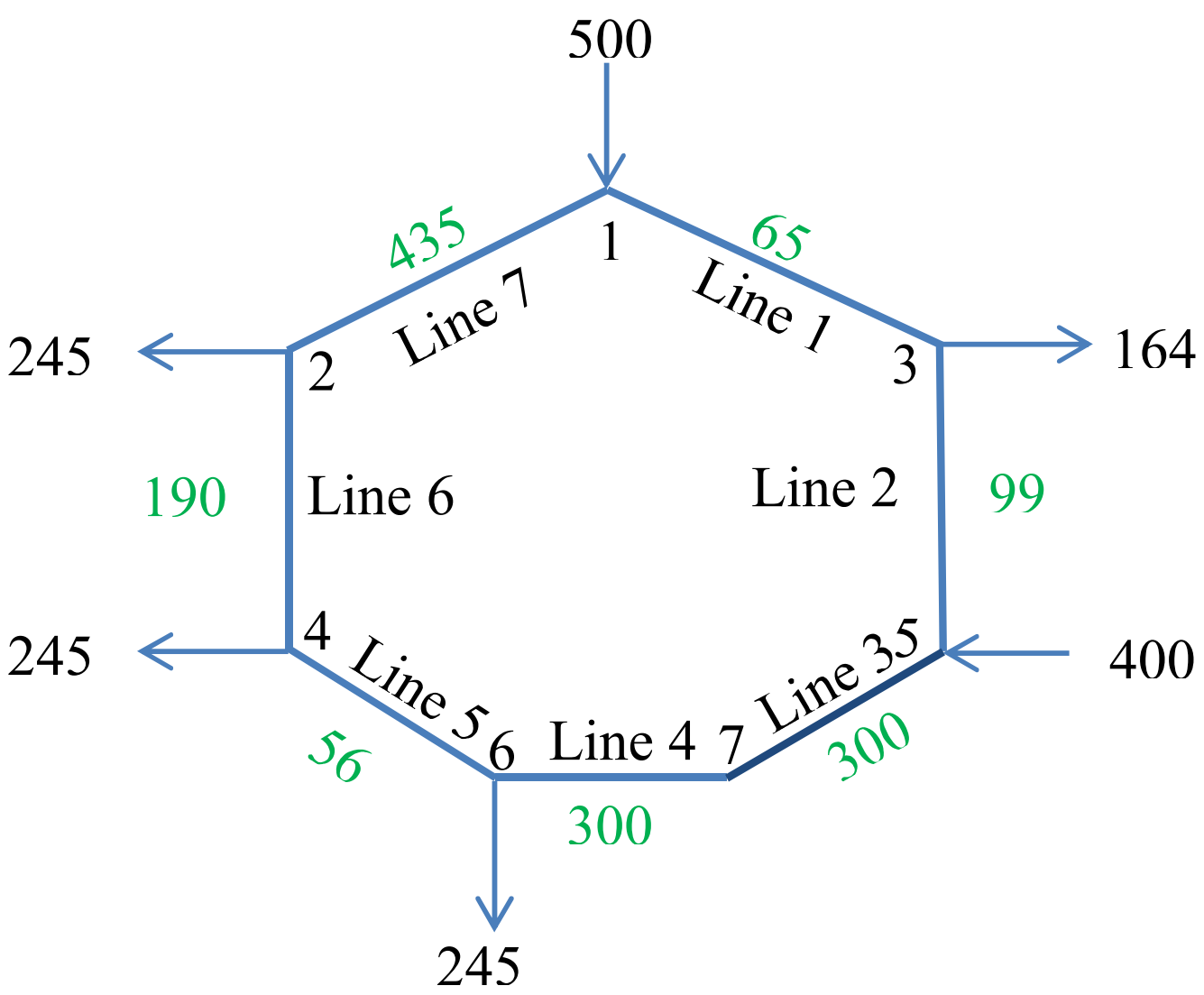

Figure 3-3 Generator 2 dispatched to 400MW

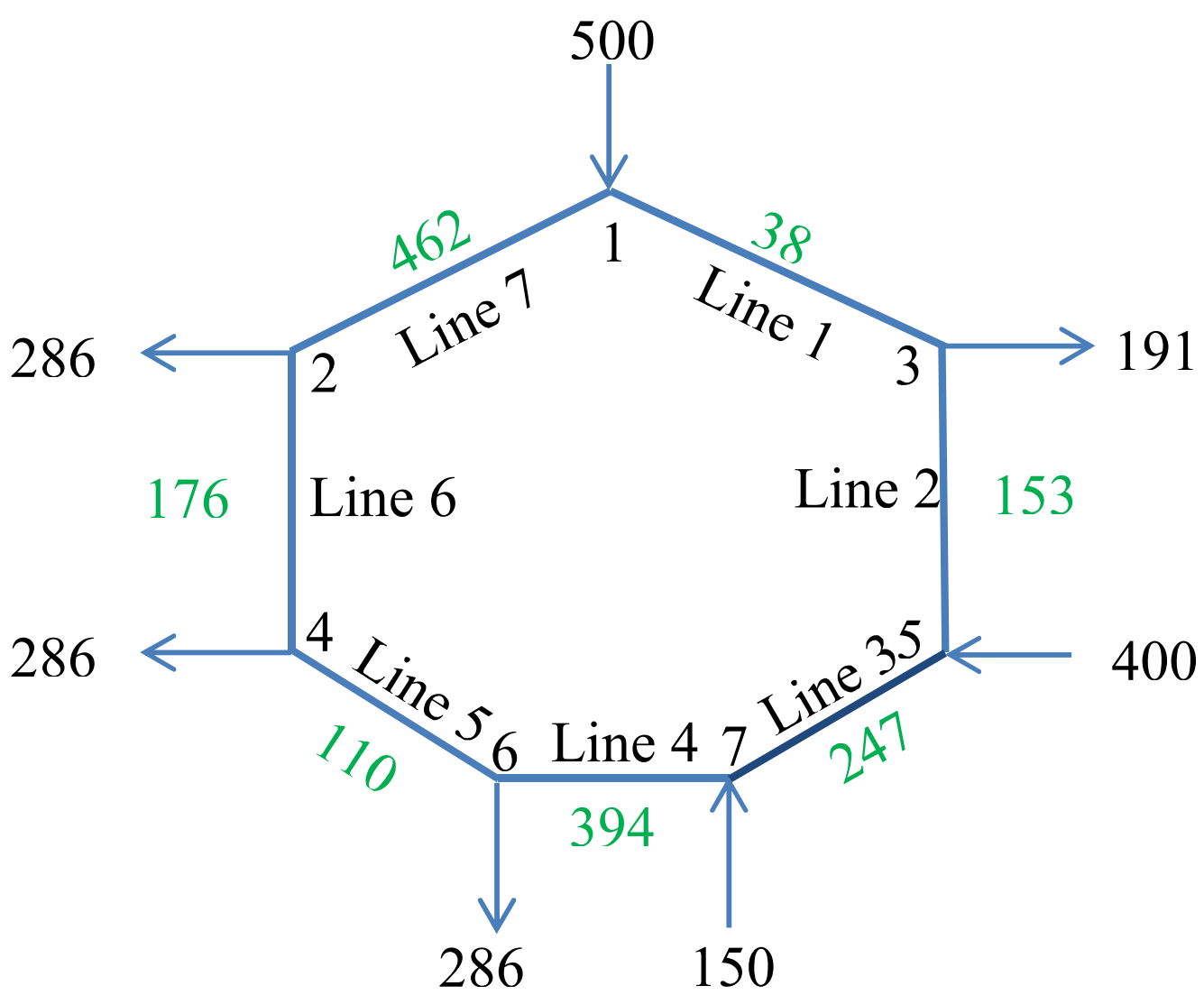

Figure 3-4 Generator 3 dispatched to 150MW

**Step 4:**

Load requirements are not yet fulfilled. However, we can't increase generation of generator 2. Therefore, we will now dispatch generator 3. We can increase the output of generator 3 to its maximum value of 150 MW without violating the flow rate constraints as shown in Figure 3-4.

**Step 5:**

Now we increase the output of generator 2 to 483.33 MW while decreasing the output of generator 3 to 116.67 and check if the line flow rate constraints are violated. As shown in the figure below the line flow constraints are not violated and all the load requirements are fulfilled. Note that line 3 in Figure 3-5, shown in yellow is fully congested is this scheme. Table 3-2 shows optimal dispatch results.

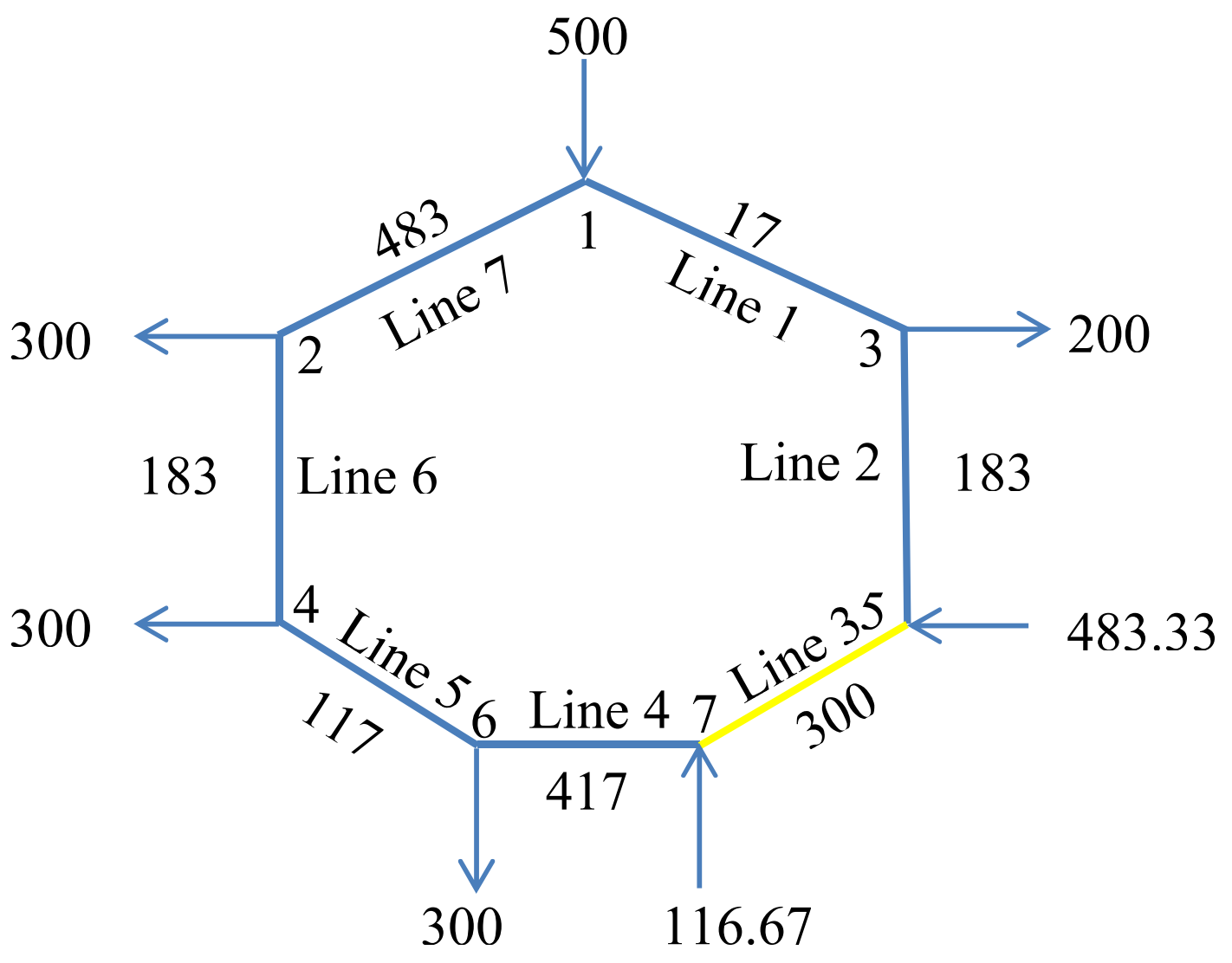

Figure 3-5 Generator 2 and 3 dispatched to 483.33MW and 116.67 respectively (Optimal Solution)

| Generator Output | | |
|---|---|---|
| $P_{G1}$ | $P_{G2}$ | $P_{G3}$ |
| 500 | 483.33 | 116.67 |

Table 3-2 Economic Dispatch using logical method when load = 1100 MW (time=17)

### 3.1.3 Nodal Prices using Logical Method

For marginal generating units the nodal prices are equal to bidding prices. Therefore, nodal prices at nodes 5 and 6 are 20 $/MW and 50 $/MW respectively.

If no line was congested the nodal prices will be equal to a clearing price of 10 $/MW. However, line 3 is fully loaded and load can't be increased using this line as shown in yellow color in Figure 3-8.

Nodal price is defined as the cost of providing an additional 1 MW of power at the node. Using this definition, I will calculate the nodal price at node 3. The additional MW of power can be provided by the marginal generation units. To provide an additional 1 MW at node 3 we need to change the output of G2 at node 5 and the output of generator G3 at node 7 so that change in flow rate at line 3 remains constant. Suppose we increase the output of generator G2 and G3 by $\Delta P_{G2}$ and $\Delta P_{G3}$ respectively. Figure 3-6 and Figure 3-7 shows line flows due to $\Delta P_{G2}$ and $\Delta P_{G3}$ respectively. The change in flow rate at line 3 will be $5/7\,\Delta P_{G3} - 1/7\,\Delta P_{G2}$. The change in flow rate at line 3 should be zero so that flow rate does not exceed its limit.

$$\frac{5}{7}\Delta P_{G3} - \frac{1}{7}\Delta P_{G2} = 0 \tag{3-1}$$

Therefore,

$$5\Delta P_{G3} - \Delta P_{G2} = 0 \tag{3-2}$$

The total change in power required is 1 MW. Therefore,

$$\Delta P_{G2} + \Delta P_{G3} = 1 \tag{3-3}$$

Solving (3-2) and (3-3) we get,

$$\begin{aligned}\Delta P_{G2} &= 0.8333 \\ \Delta P_{G3} &= 0.1667\end{aligned} \tag{3-4}$$

Therefore, the cost of increasing power by 1MW at node 3 is given by,

$$20\Delta P_{G2} + 50\Delta P_{G3} = 25 \tag{3-5}$$

Therefore, the nodal price at node 3 is 25$/MW.

We can use the same procedure to find nodal prices at each node. Figure 3-8 shows the nodal prices in red calculated using logical method.

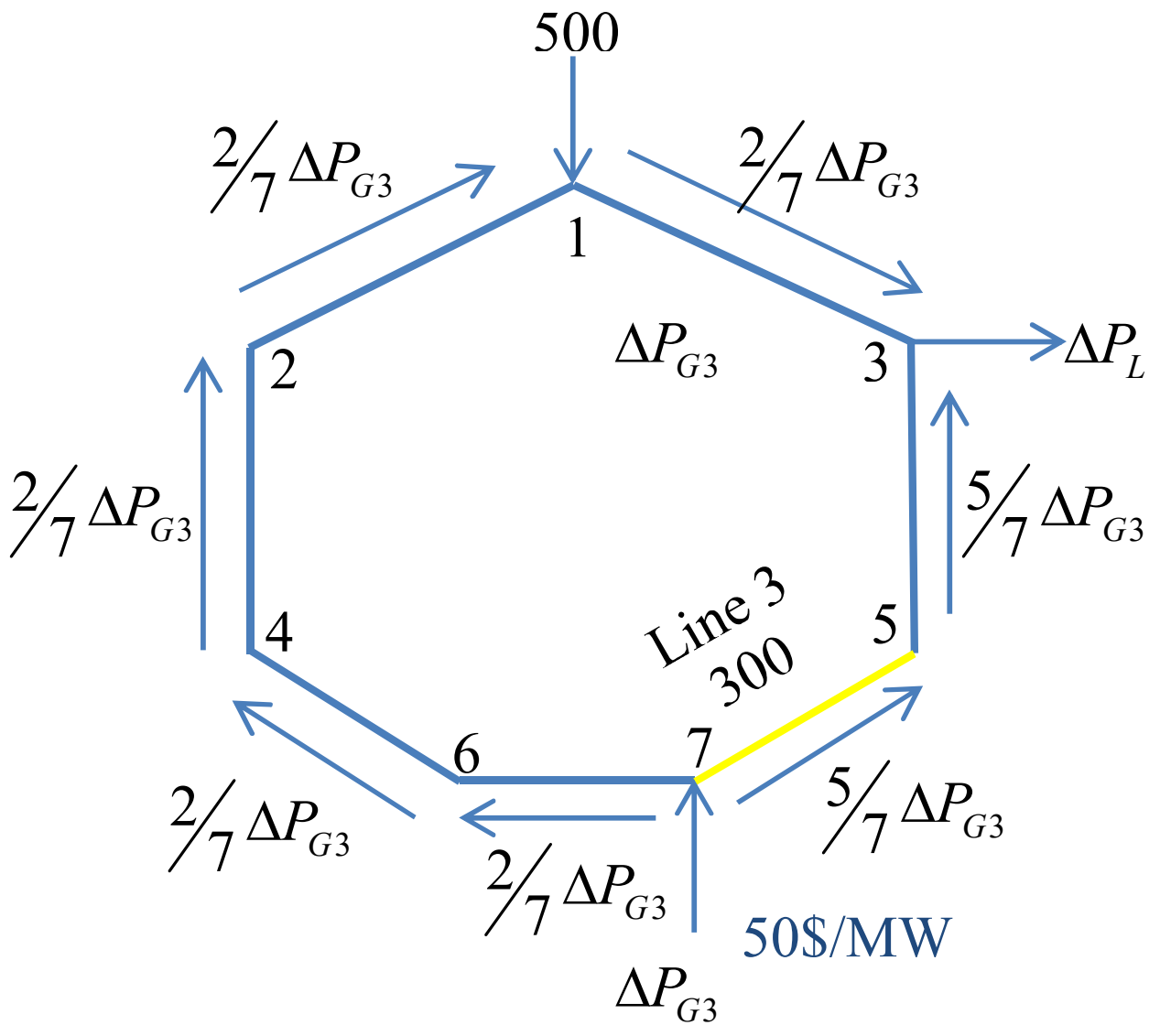

Figure 3-7 Line flows due to $\Delta P_{G3}$

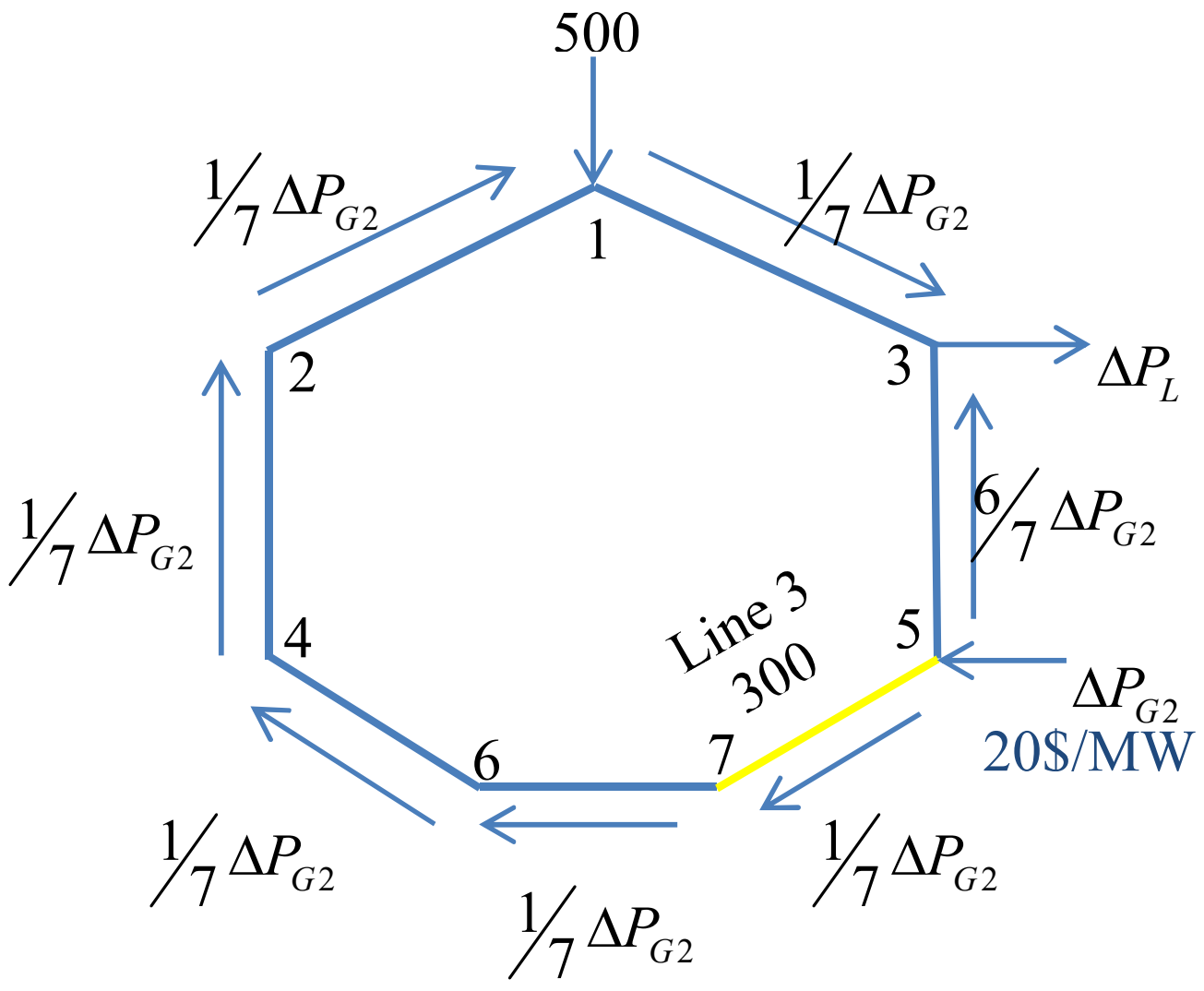

Figure 3-6 Line flows due to $\Delta P_{G2}$

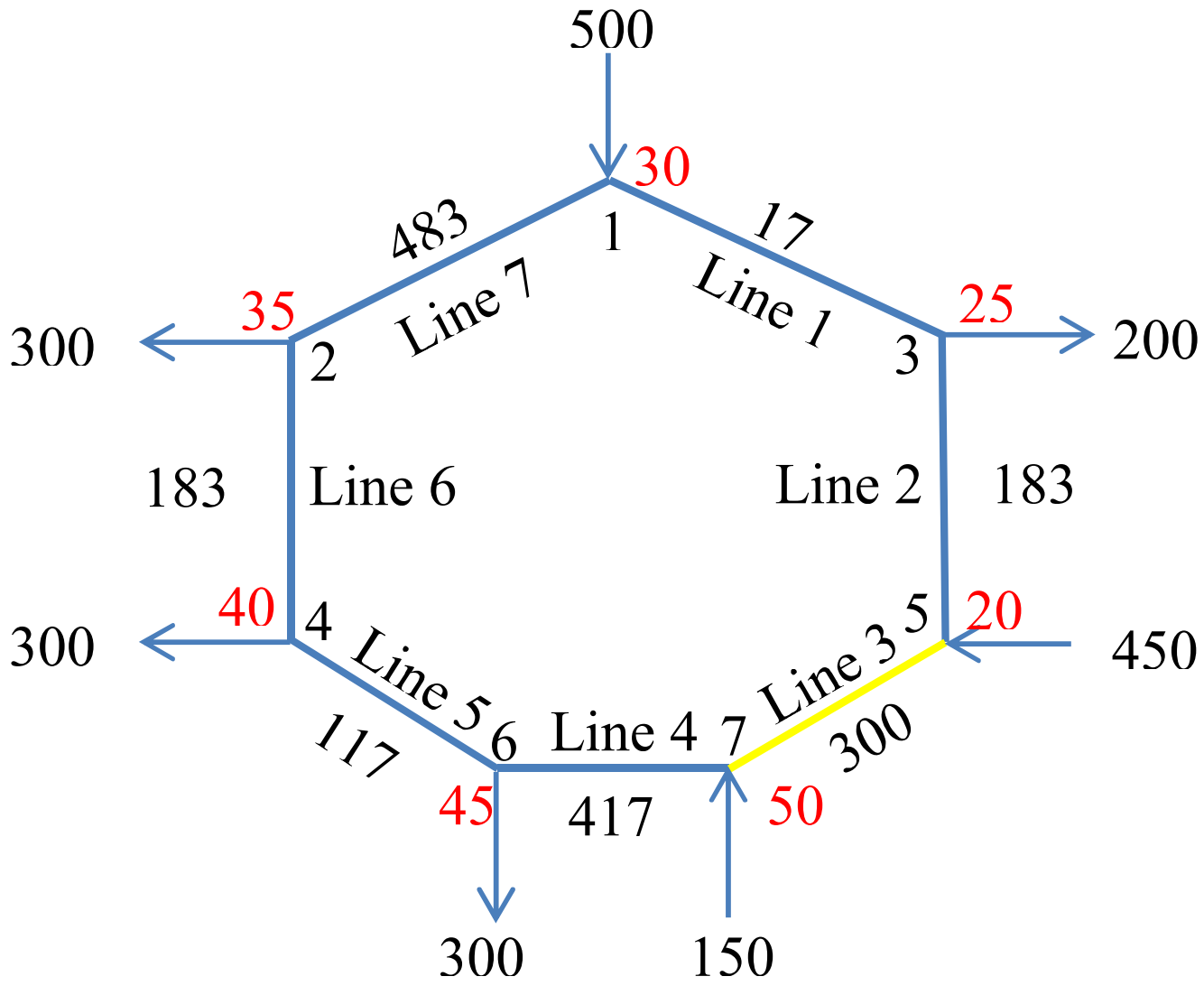

Figure 3-8 Nodal prices using Logical Method

## 3.2 Results of Day-Ahead Market Problem

Figure 3-9 and Figure 3-10 shows the hourly day-ahead nodal prices at each bus. We can clearly see that solar generator has decreased the average nodal prices. This means cheap availability of electricity for consumers. At peak hours the nodal prices for most nodes has also decreased as shown in Table 3-3.

However, nodal price for bus 7 remains the same during peak hours. This is because of congestion in line 3. Generator 3 at bus 7 is marginal generator and nodal price is equal to its bidding price (50 $/MW). Therefore, customers connected to this bus will not have any benefit of cheap solar energy. Line congestion can hinder the benefits of cheap electricity from reaching the customers. Therefore, the topology of the power grid and line constraints should be carefully considered when solar power plants are set up so that all consumers can enjoy maximum benefit from cheap electricity.

| | **Nodal Prices at peak hours t=13 to 21 ($/MW)** | | | | | | |
|---|---|---|---|---|---|---|---|
| | $\rho_1$ | $\rho_2$ | $\rho_3$ | $\rho_4$ | $\rho_5$ | $\rho_6$ | $\rho_7$ |
| **No Solar Generator** | 30 | 35 | 25 | 40 | 20 | 45 | 50 |
| **Using Solar Generator Gx** | 20 | 27 | 12 | 35 | 5 | 42 | 50 |

Table 3-3 Comparing nodal prices for conventional and renewable sources

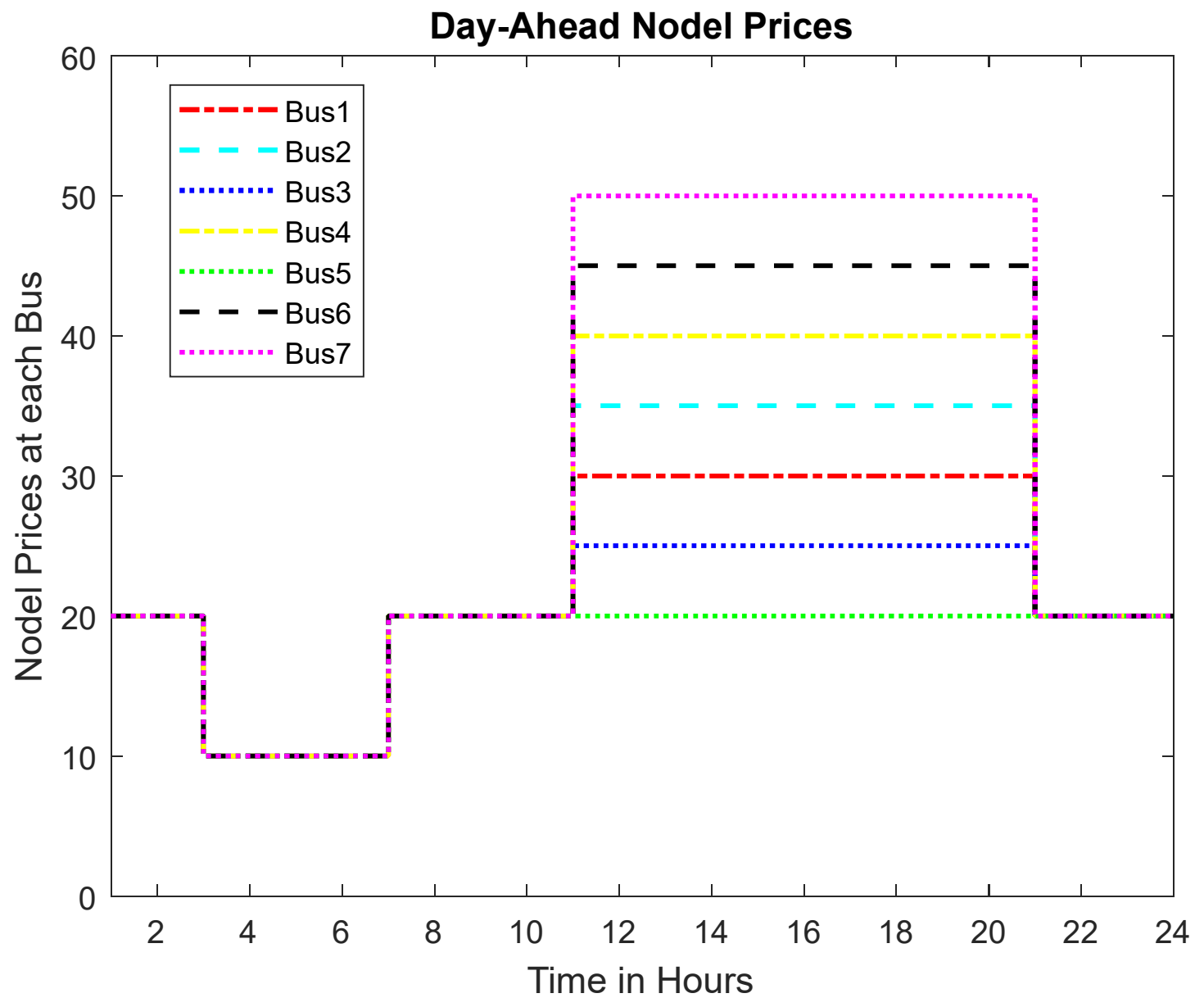

Figure 3-9 Hourly Day-Ahead nodal prices at each bus

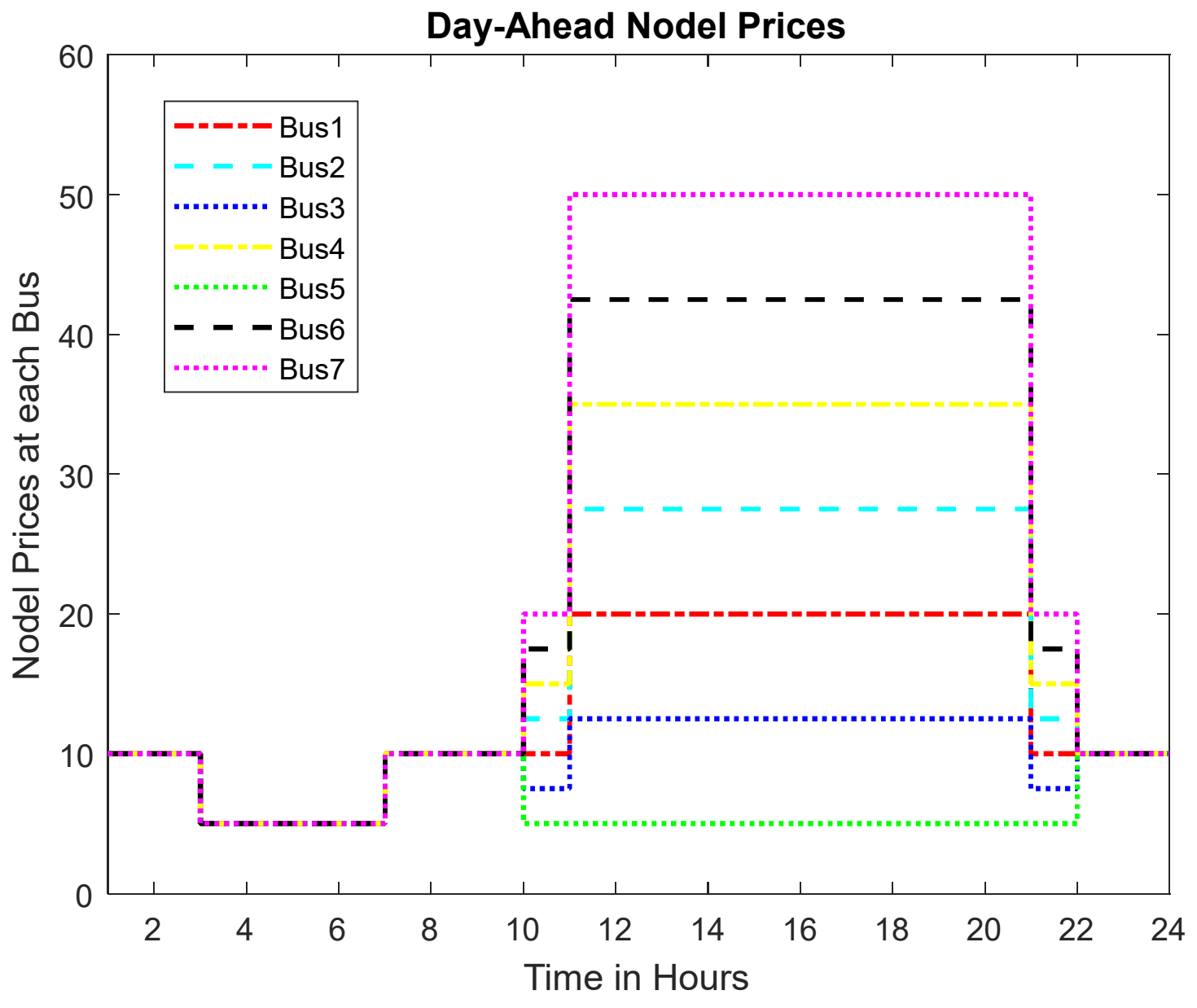

Figure 3-10 Hourly Day-Ahead nodal prices at each bus when solar generator Gx is used

# 4 Conclusion

Renewable energy can decrease the overall cost of electricity. Increasing renewable energy sources will decrease electricity spot prices and consumers can enjoy cheap electricity. However, topology of power grid must be carefully analyzed before setting up solar and other renewable energy power plants. As shown in this report, line congestion can cause some nodes to retain the same high nodal prices. This is especially the case during peak hours. Therefore, new lines need to be installed or the capacity of old ones should be increased so that the loss of benefits due to line congestion can be avoided.

# Appendix

## Matlab Code

```
Nunits = 7;
Horizon = 24;
%Bidding Price
%p=[10 0 0 0 20 0 50];%Conventional thermal power plant
p=[10 0 0 0 5 0 50];%Using solar power plant
%Startup cost of Generator1=500
%Startup cost of Generator2=300
%Startup cost of Generator3=50
%Scost = [500 0 0 0 100 0 50];%Conventional thermal power plant
Scost = [500 0 0 0 0 0 50];%Using solar power plant
%Forcasted values of Load
Pforecast = (1100/100).*[60 50 40 37 40 45 56 63 65 75 85 90 93 96 98 100 100
90 87 85 70 68 65 60];

x=[0.05 0.05 0.05 0.05 0.05 0.05 0.05];
X=[-1/x(1) 1/x(1) 0 0 0 0 0;
-1/x(2) 0 1/x(2) 0 0 0 0;
0 1/x(3) 0 -1/x(3) 0 0 0;
0 0 1/x(4) 0 -1/x(4) 0 0;
0 0 0 1/x(5) 0 -1/x(5) 0;
0 0 0 0 1/x(6) 0 -1/x(6);
0 0 0 0 0 1/x(7) -1/x(7)];

B=[40 -20 -20 0 0 0 0;
-20 40 0 -20 0 0 0;
-20 0 40 0 -20 0 0;
0 -20 0 40 0 -20 0;
0 0 -20 0 40 0 -20;
0 0 0 -20 0 40 -20;
0 0 0 0 -20 -20 40];

Bx=B(1:6,1:6);
T=X*[inv(Bx) zeros(6,1);zeros(1,7)];
Fl=[500 800 500 300 800 300 800]';

%Minimum/Maximum Output Power of Generators
%Conventional thermal power plant
%Pmin=[20 0 0 0 10 0 0]';
%Pmax=[500 0 0 0 500 0 150]';
%Using solar power plant
Pmin=[20 0 0 0 0 0 0]';
Pmax=[500 0 0 0 500 0 150]';

onoff = binvar(Nunits,Horizon,'full');
P      = sdpvar(Nunits,Horizon,'full');
shutdown = binvar(Nunits,Horizon,'full');

Constraints = [];
for k = 1:Horizon
```

```
  Constraints = [Constraints, onoff(:,k).*Pmin <= P(:,k) <=
onoff(:,k).*Pmax];
end

for k = 1:Horizon
    L=[0 3/11 2/11 3/11 0 3/11 0]'.*Pforecast(k);
    Blin=Fl+T*L;
  Constraints = [Constraints, T*P(:,k)<=Blin];
end

for k = 1:Horizon
  Constraints = [Constraints, sum(P(:,k)) == Pforecast(k)];
end

Objective = 0;
for k = 1:Horizon
  Objective = Objective + p*P(:,k) + Scost*shutdown(:,k);
end

for k = 2:Horizon
 for unit = 1:Nunits
  Constraints = [Constraints, shutdown(unit,k) >= (onoff(unit,k)-
onoff(unit,k-1))];
 end
end

ops = sdpsettings('verbose',1,'debug',1);
optimize(Constraints,Objective,ops)
stairs(value(P(1,:)));
hold on
stairs(value(P(5,:)));
stairs(value(P(7,:)));
stairs(Pforecast);
legend('Generator 1','Generator 2','Generator 3','Load');
xlabel('Time in Hours') % x-axis label
ylabel('Generator Output (MW)') % y-axis label
title('Generator Scheduling')
axis([1 24 0 1200])

%Plot of overall social production cost
figure(2)
stairs(p*value(P));
xlabel('Time in Hours') % x-axis label
ylabel('Social Production Cost (Dollars)') % y-axis label
title('Social Production Cost')

%generators' turn on periods
%generator 1
X1=value(P(1,:));
X1(find(value(P(1,:))))=1;
%generator 2
X2=value(P(5,:));
X2(find(value(P(5,:))))=1;
%generator 3
X3=value(P(7,:));
```

```
X3(find(value(P(7,:))))=1;
figure(3)
stairs(X1, '-.r','LineWidth',1.5)
hold on
stairs(X2, '--b','LineWidth',1.5)
hold on
stairs(X3, ':y','LineWidth',1.5)
legend('Generator 1','Generator 2','Generator 3');
axis([1 24 0 1.5])
xlabel('Time in Hours') % x-axis label
ylabel('ON/OFF Status of Generators') % y-axis label
title('Generator ON/OFF Status during 24 hour period')

%Solve the generator dispatch problem at each time interval
%to find nodal prices at each time interal
price=ones(7,24);%nodal prices at each time interval
for k = 1:Horizon
    L1=[0 3/11 2/11 3/11 0 3/11 0]'.*Pforecast(k);
    P1=sdpvar(Nunits,1);
    s=[X1(k) 0 0 0 X2(k) 0 X3(k)]';
    Pmin1=[20 0 0 0 10 0 0]'.*s;
    Pmax1=[500 0 0 0 500 0 150]'.*s;
    Objective1=p*P1;
    Constraints1=[T*(P1-L1)<=Fl -T*(P1-L1)<=Fl ones(1,7)*P1==ones(1,7)*L1
Pmin1<=P1 P1<=Pmax1];
    ops = sdpsettings('verbose',1,'debug',1,'saveduals',1);
    optimize(Constraints1,Objective1,ops);
    value(P1)
    %Line Flows
    value(T*(P1-L1))
    %Nodal Prices
    price(:,k)=-dual(Constraints1(3))-T'*dual(Constraints1(1))
end

%plotting nodal prices
figure(4)
stairs(price(1,:)','-.r','LineWidth',1.5)
hold on
stairs(price(2,:)','--c','LineWidth',1.5)
stairs(price(3,:)',':b','LineWidth',1.5)
stairs(price(4,:)','-.y','LineWidth',1.5)
stairs(price(5,:)',':g','LineWidth',1.5)
stairs(price(6,:)','--k','LineWidth',1.5)
stairs(price(7,:)',':m','LineWidth',1.5)
legend('Bus1','Bus2','Bus3','Bus4','Bus5','Bus6','Bus7');
axis([1 24 0 60])
xlabel('Time in Hours') % x-axis label
ylabel('Nodel Prices at each Bus') % y-axis label
title('Day-Ahead Nodel Prices')
```